\documentclass{article}  
\usepackage{bigsky2009}
\usepackage{graphicx}
\usepackage{subfigure}
\usepackage{mcite}
\frompage{000} \topage{000}                                              
\def\rightmark{Heavy Flavor Electron Correlations at PHENIX}
\def\leftmark{Anne Sickles et al.}
 
\title{Correlations of Electrons from Heavy Flavor Decay with Hadrons at PHENIX} 
\authors{
{Anne Sickles$^1$ for the PHENIX Collaboration  %
}\\[2.812mm]
{\normalsize
\hspace*{-8pt}$^1$ Brookhaven National Laboratory, \\ 
Upton, NY 11973\\[0.2ex] 
}}
 
\abstract{
One unexpected recent result from heavy-ion collisions is the
large suppression and elliptic flow of electrons from heavy flavor
decay.  Further measurements of properties of electrons from heavy
flavor decay are crucial to understanding the origin of this
suppression  and its implications for the properties of the hot matter
produced at RHIC. 
Two particle correlations have been used extensively to study
the propagation of hard partons through the 
produced matter in heavy-ion collisions.  
Measurements in p+p collisions are important both as
reference for heavy ion measurements and to study heavy
flavor production and fragmentation in the vacuum.  Preliminary
results of correlations between electrons from heavy flavor
decay with charged hadrons from p+p collisions are shown.}

\PACS{specifications see, e.g.\ {\tt http://www.aip.org/pacs/}}
 
\begin{document}
 
\maketitle
\setcounter{page}{1}

\section{Introduction}\label{intro}
Two particle azimuthal correlations at the Relativistic Heavy Ion Collider have
been shown to be enormously powerful to study the propagation of jet partons through
the hot nuclear matter produced in nucleus-nucleus collisions.  The same measurements
in proton-proton collisions provide an essential baseline for interpreting heavy ion results
and have also been interesting in understanding jet fragmentation~\cite{ppg029}.

As of now these jet-induced correlations have only involved light hadrons.  However,
with recent results showing the suppression of electrons from the decay of charm and
bottom mesons at a level similar to that of $\pi^0$ ~\cite{ppg066} there is great
interest in looking at the correlations of these electrons with other hadrons in the
event to study their jet structure and how it differs from that observed in proton-proton
collisions and also from the jet structure of light hadron jet correlations.  The proton-proton
collisions are especially necessary for heavy flavor correlations because the measured
electron is a decay product, thus hadrons from the heavy meson 
decay contribute to the correlations
in addition to other fragmentation products.  
 
\section{Analysis Method}
Studies of electrons from heavy flavor decay are complicated by the large background
of electrons from light meson decay and photon conversions.  In this analysis we statistically
subtract the correlations from these photonic sources using a method similar to that
used to measure direct photon-hadron correlations~\cite{ppg090}.  What is measured is
the inclusive electron-hadron conditional hadron yields per trigger
electron, $Y_{e_{inc}-h}$.  This is a weighted average
of conditional yields hadrons associated with electrons from
 electrons from heavy flavor decay and the conditional yield of hadrons
associated with electrons from photonic sources:
\begin{equation}
Y_{e_{incl}-h} = \frac{N_{e_{HF}} Y_{e_{HF}-h} + N_{e_{phot}} Y_{e_{phot}-h}}{N_{e_{HF}} + N_{e_{phot}}}
\end{equation}

$Y_{e_{HF}-h}$ is  then given by:
\begin{equation}
Y_{e_{HF}-h} = \frac{(R_{HF}+1)Y_{e_{inc}-h} - Y_{e_{phot}-h}}{R_{HF}}
\label{sub_eq}
\end{equation}
where $R_{HF}\equiv \frac{N_{e_{HF}}}{N_{e_{phot}}}$ and is taken from the published
PHENIX measurement~\cite{ppg065} for the electron $p_T$ bins used in this analysis.

\begin{figure}[t]
\centering
\includegraphics[width=\textwidth]{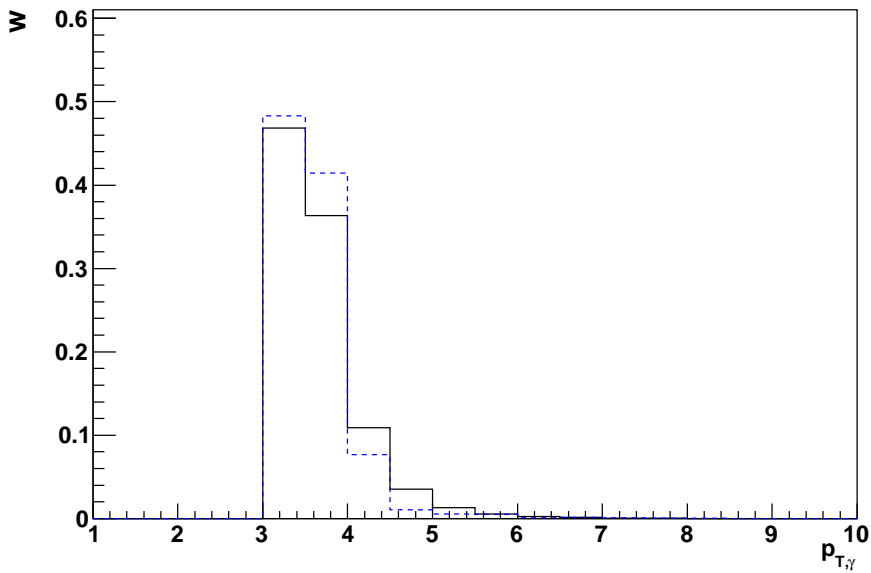}
\caption{$w$ for electrons with 3.0$<p_T<$3.5GeV/c from the method using Dalitz decays
(black solid line) and photon conversions (blue dashed line) as a function of the photon
$p_T$.}
\label{weights}
\end{figure}

The remaining unknown is $Y_{e_{phot}-h}$.  In the electron $p_T$ range of interest here,
1.5$<p_T<$4.5GeV/c the photonic electrons are dominated by $\pi^0$ Dalitz decay and photon
conversions~\cite{ppg065}, where the primary source of the photons which convert
is $\pi^0\to\gamma\gamma$ decay.  
We then measure inclusive photon-hadron correlations, where the inclusive photons are 
dominately from $\pi^0$ decay.  Simulations are used to construct $Y_{e_{phot}-h}$
from $Y_{\gamma_{inc}-h}$:
\begin{equation}
Y_{e_{phot}-h}(p_{T,i}) = \sum_j\ w_i(p_{T,j})\ Y_{\gamma_{inc}-h}(p_{T,j})
\end{equation}
where each $i$ and $j$ represents a bin in electron or photon $p_T$, respectively, 0.5GeV/c
wide.

Two methods are used to determine the $w_i(p_{T,j})$ coefficients.  In the first method the measured
inclusive photon spectrum in p+p collisions is input into a GEANT based simulation of the 
PHENIX detector.  The same electron identification 
cuts as in the real data analysis are then applied to reconstructed
conversion electrons and the relationship between the input photon $p_T$ and the reconstructed
conversion electron $p_T$ determines $w$.  In the second method, the measured $\pi^0$
spectrum from~\cite{ppg024} is taken as input to a Monte Carlo simulation which decays the
$\pi^0$s via their Dalitz decay.  The relationship between the
intermediate off-shell photon and the resulting electron are used in the same manner as in 
the first method to determine $w$.  The resulting weights are shown in Fig.~\ref{weights}
for a single electron $p_T$ selection.  
The difference between the two methods is small.  The $\pi^0$ spectrum falls steeply
with $p_T$ and thus the measured photonic electrons at a given $p_T$ are dominated by those
carrying a large fraction of the $\pi^0$ $p_T$ regardless of whether the electron comes from a 
conversion or a Dalitz decay.

The measured conditional yields are corrected for the hadron acceptance and 
reconstruction efficiency by comparing the raw measured
hadron $p_T$ spectrum to the corrected spectrum in Ref.~\cite{ppg050}.  The non-uniform
azimuthal acceptance in PHENIX is corrected for by using mixed pairs. 
The jet functions are obtained by subtracting the combinatoric background
by the zero yield at minimum (ZYAM)~\cite{ppg083} method.

The electrons were identified as in Refs.~\cite{ppg065,ppg094}.
The systematic errors are from the different methods used to construct $Y_{e_{phot}-h}$,
the statistical error on the determination of the ZYAM level which moves all points by the same
amount in 
a given $p_{T,e}$ and $p_{T,h}$ combination.  For $Y_{e_{HF}-h}$ there is an additional uncertainty
from the $R_{HF}$ value used to subtract the jet functions. These systematic
errors are shown the grey boxes in the plots that follow. The systematic error on the efficiency
correction is 10\% and moves all conditional yield measurements together and is not shown on the plots.

\section{Results}
\begin{figure}[t]
\centering
\includegraphics[width=\textwidth]{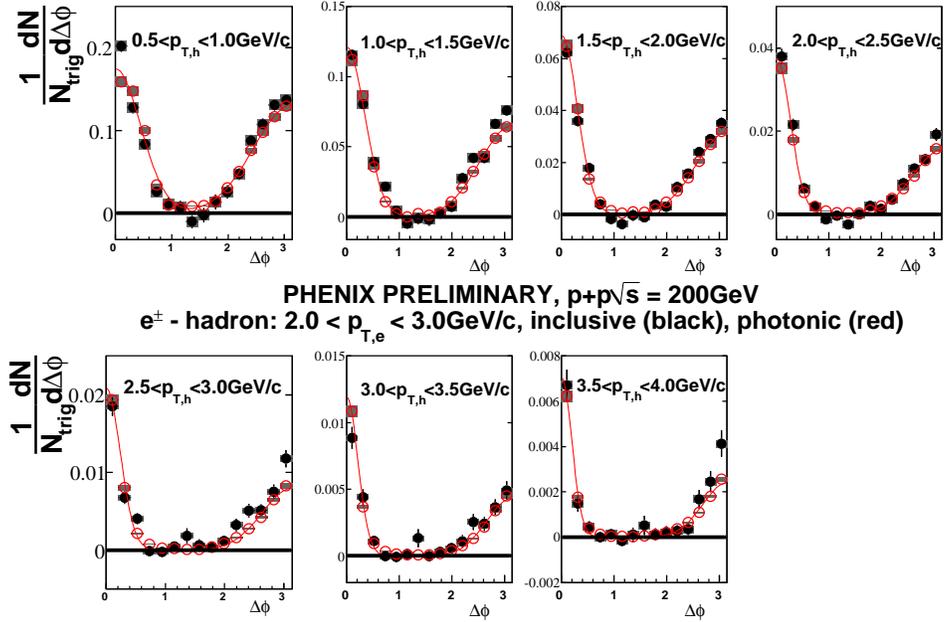}
\caption{Efficiency corrected $e_{inc}-h$ jet functions
 (solid black) and ${e_{phot}}-h$ jet functions (hollow red) 
conditional yields for 
electrons with 2.0$<p_T<$3.0GeV/c and seven hadron $p_T$ bins from 0.5-4.5GeV/c.}
\label{electron_dphi}
\end{figure}

\begin{figure}[t]
\centering
\includegraphics[width=\textwidth]{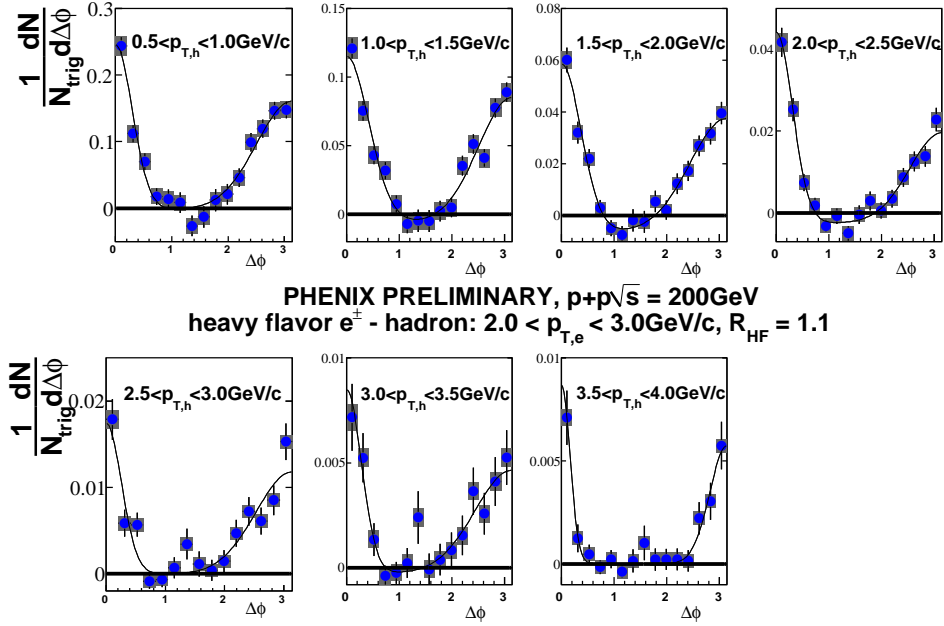}
\caption{$Y_{e_{HF}}-h$ conditional yields following the subtraction in Eq.~\ref{sub_eq} for
2.0$<p_{T,e}<$3.0GeV/c and seven hadron $p_T$ bins from 0.5-4.5GeV/c.  In this $p_T$ bin $R_{HF}=$1.1. }
\label{hf_dphi}
\end{figure}

Fig.~\ref{electron_dphi} shows the corrected jet functions for inclusive electrons with
2.0$<p_T<$3.0GeV/c and seven hadron $p_T$ bins.  The constructed photonic electron jet functions
are shown in the hollow points.  The fits shown are to near and away side Gaussians with a flat
background.  The resulting $e_{HF}-h$ jet functions are extracted using Eq.~\ref{sub_eq} and are
shown in Fig.~\ref{hf_dphi}.  

\begin{figure}[t]
\centering
\includegraphics[width=\textwidth]{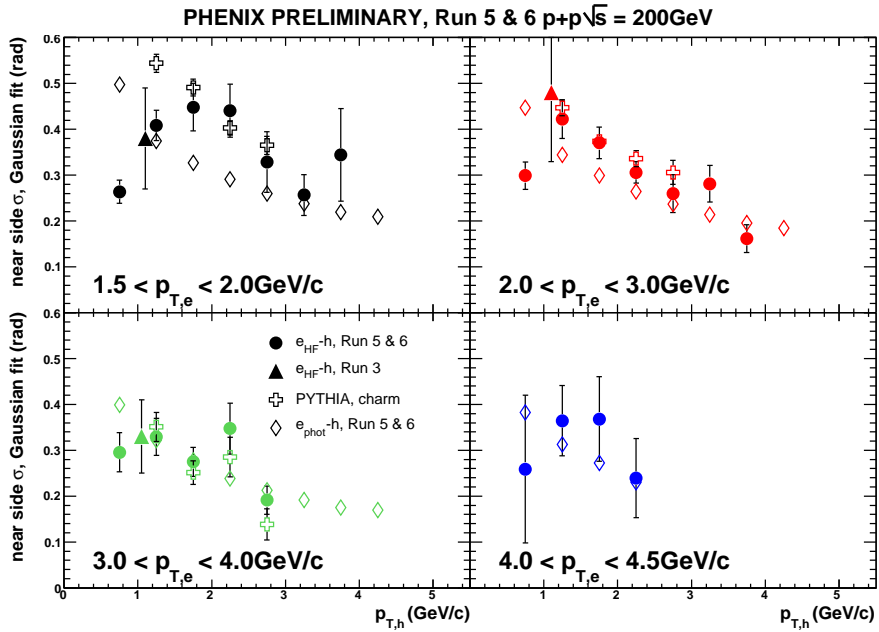}
\caption{Gaussian widths as a function of $p_T,h$ for four electron $p_T$ bins.  Circles
show the $e_{HF}-h$ widths and open diamonds show $e_{phot}-h$ widths.  Pluses show
$e_{HF}-h$ widths from PYTHIA and the triangles show an earlier PHENIX measurement
of $e_{HF}-h$ widths with lower statistics and 1.0$<p_T<$4.0GeV/c.}
\label{widths}
\end{figure}

In heavy quark fragmentation most of the quark momentum is carried by the heavy 
meson~\cite{cleoc,*bellec,alephb,*opalb,*sldb}, thus
the near side $e_{HF}-h$ is expected to be dominated by pairs in which both particles
come from the heavy meson decay.  Due to the decay kinematics, we then expect 
the Gaussian widths of the near side to be wider for $e_{HF}-h$ than $e_{inc}-h$.  The widths
from both categories of electrons are shown in Fig.~\ref{widths}.  The measured 
near side widths for
$e_{HF}-h$ are wider than $e_{phot}-h$ and are in agreement with widths from charm production
in PYTHIA~\cite{pythia}.

\begin{figure}
\centering
\subfigure{
\includegraphics[width=0.45\textwidth]{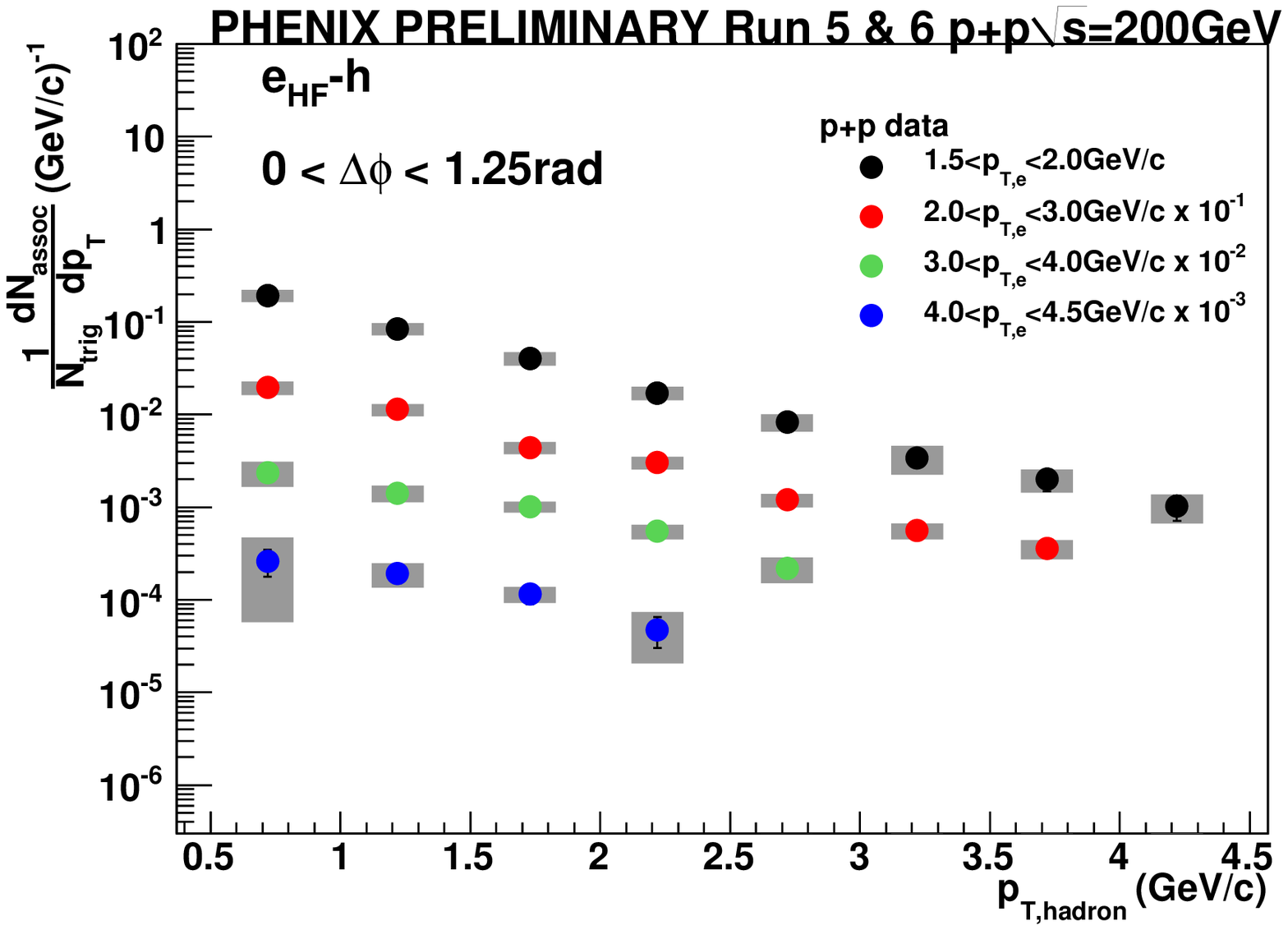}}
\subfigure{
\includegraphics[width=0.45\textwidth]{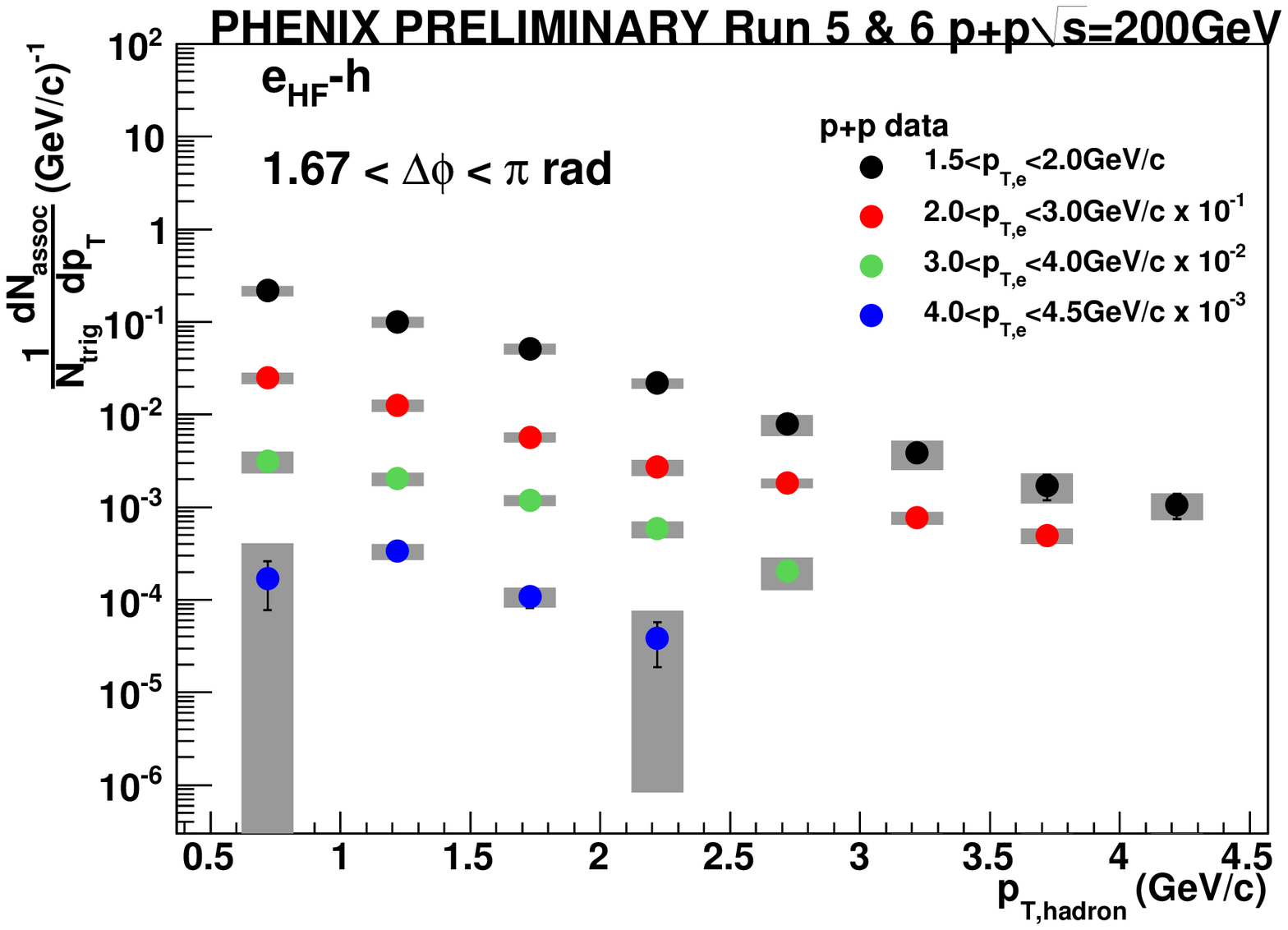}}
\caption{Near (left) and away (right) side conditional yields for $e_{HF}-h$ correlations
for four electron $p_T$ bins and eight hadron $p_T$ bins. }
\label{yield_figs}
\end{figure}

The jet functions are then integrated to get near and away side conditional yields of hadrons 
per trigger heavy flavor electron.  
The near side integration range is 0$<\Delta\phi<$1.25~rad and the away 
side integration range is 1.67$<\Delta\phi<\pi$~rad.  
The conditional yields are shown in Fig.~\ref{yield_figs}.

\begin{figure}
\centering
\subfigure{
\includegraphics[width=0.45\textwidth]{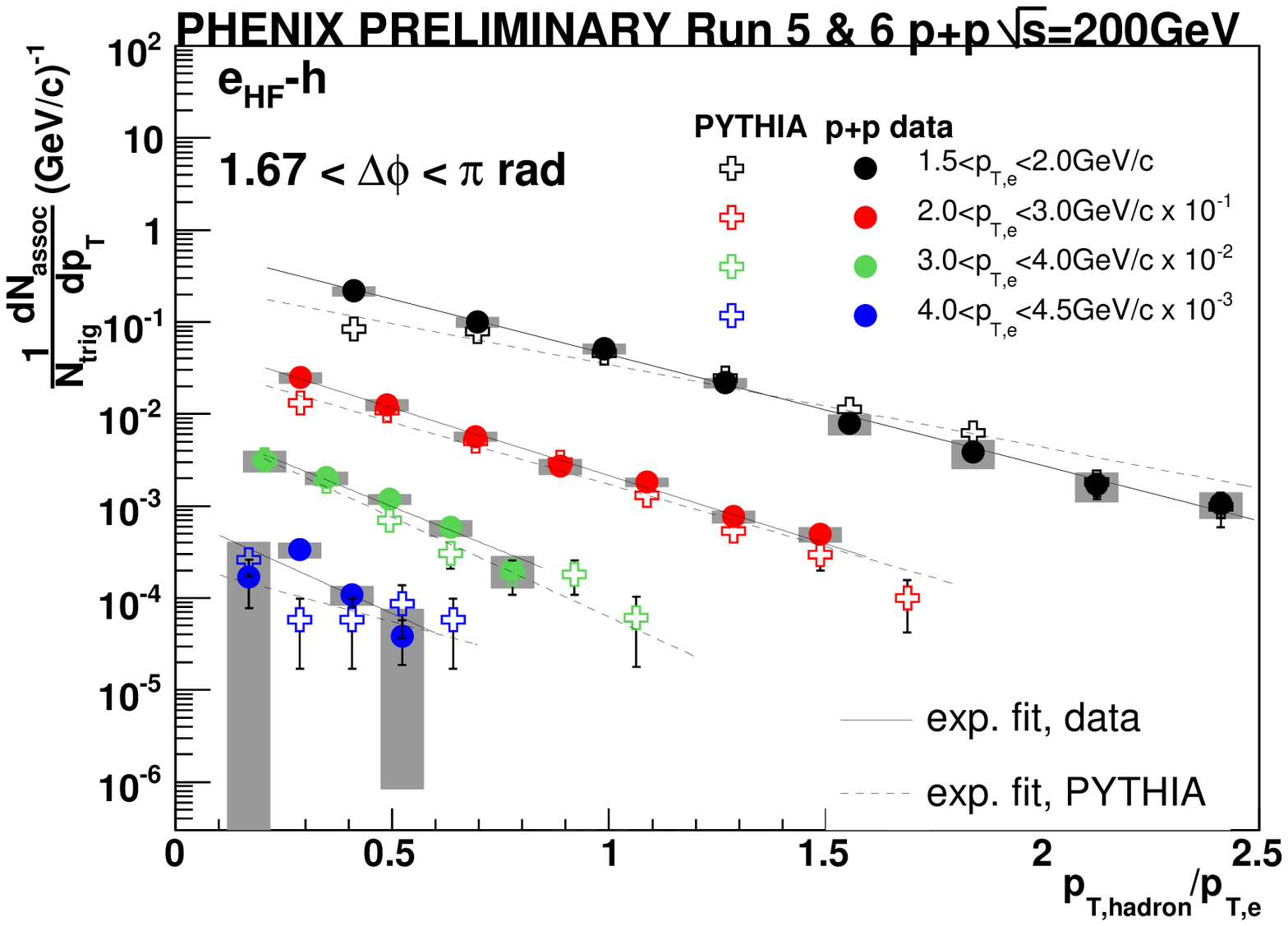}}
\subfigure{
\includegraphics[width=0.45\textwidth]{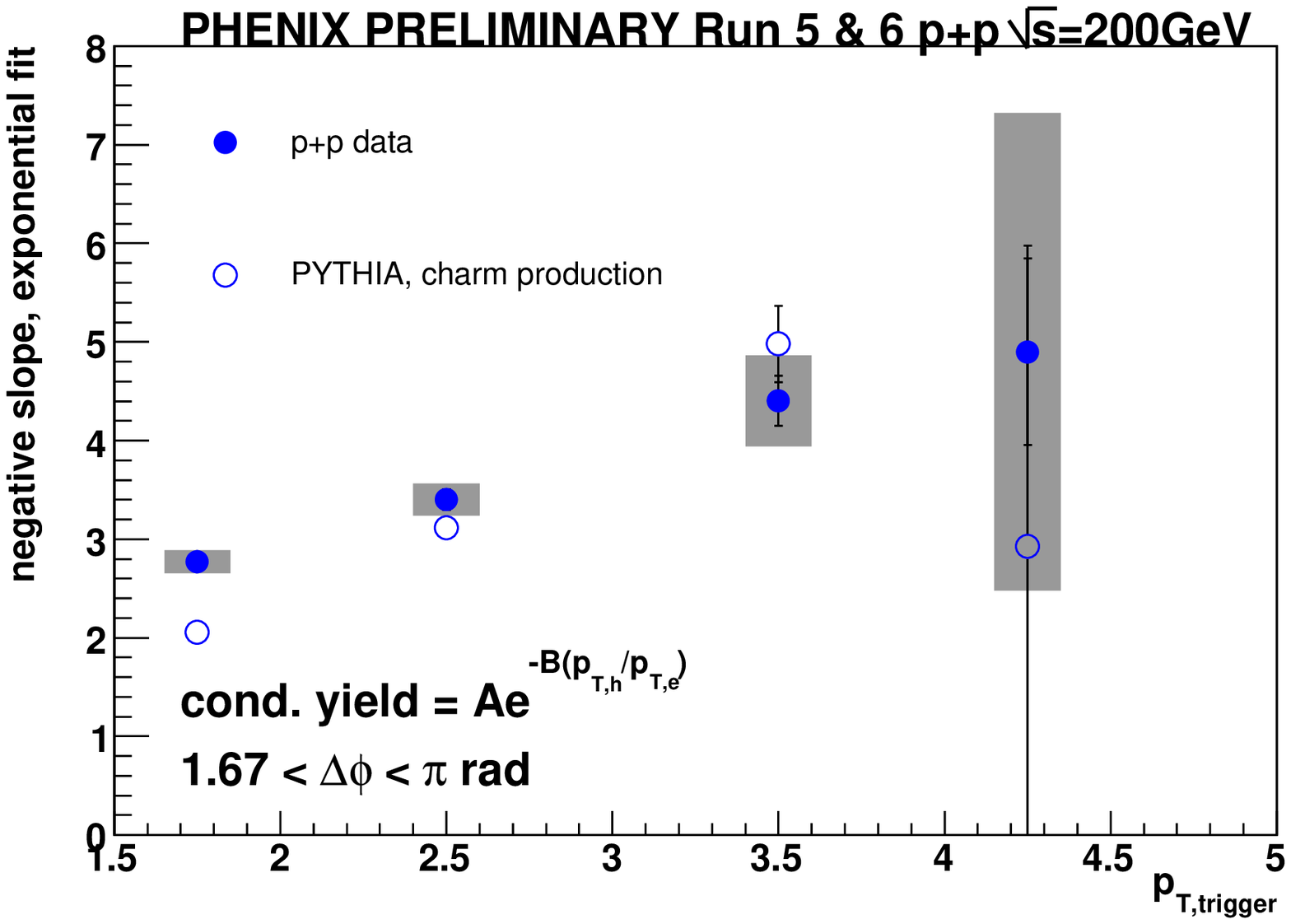}}
\caption{(left)Away side conditional yield measurements as a function of 
$p_{T,hadron}/p_{T,e}$.  Fits for the data and the PYTHIA are to exponentials.
(right) Negative slopes from the data and PYTHIA  fits.}
\label{slope_fig}
\end{figure}

The left panel of Fig.~\ref{slope_fig} shows the away side conditional yields for the data 
 as a function of $p_{T,hadron}/p_{T,e}$ along with exponential fits.  
For comparisons we also show the same distributions and fits for a PYTHIA simulation of charm
production (MSEL=4).
The negative slopes are shown
from the data and PYTHIA are shown in the right panel of Fig.~\ref{slope_fig}.  The ratio 
$p_{T,hadron}/p_{T,e}$ is defined as $z_T$ for the light flavor jets.  We can compare the away side
slopes measured here to those measured in Fig. 19 of Ref.~\cite{ppg029} for $\pi^0-h$ correlations.  
At the same $p_{T,trig}$ the $e_{HF}-h$ slopes are harder than the $\pi^0-h$.  However, in both
cases the $p_{T,trig}$ is not the $p_{T,parton}$.  

Since the current measurements are at low $p_T$ where sizable contributions of electrons from 
charm quarks the PYTHIA comparisons made here are only for charm production.  In order to
make more detailed comparisons, it is necessary to include bottom quarks as well in 
addition to including next to leading order effects not included in PYTHIA.  
A further complication is that the heavy flavor quark that leads to a trigger is not necessarily 
balanced by an away side balancing heavy quark.  Ref~\cite{ivanhq} shows a leading order calculation
of the subprocess leading to mid-rapidity $D$ production and finds that in the $p_T$ range relevant
here $gg\to c\bar{c}$ contributes $\approx$20\% and the rest of the contribution is from processes
such as $cg\to cg$ or $cq(\bar{q}) \to cq(\bar{q})$.  These processes are included in next to
leading order calculations where there are three partons in the final state~\cite{hqnlo}.  
However the PYTHIA simulations here only include $gg\to c\bar{c}$.
Further
studies using NLO Monte Carlo calculations are necessary to understand the partonic contribution to the
away side jets. 

\section{Conclusions and Outlook} 
Jet physics with heavy flavor is an important new frontier in heavy ion physics. 
Heavy flavor triggered two particle correlations provide complementary information
to $\pi^0$ and direct photon triggered measurements.  However the measurement are difficult
because of the background from photon conversions and light meson Dalitz decays.  We
have shown a new method which allows the extraction of correlations triggered by electrons from
heavy flavor decay and showed preliminary results from proton-proton collisions.  
These measurements are an essential baseline for understanding heavy ion results.  These
measurements are crucial to understanding the interactions between heavy quarks and 
the hot nuclear matter produced in relativistic heavy ion collisions.  

Current measurements are limited by statistics and by the inability to separate the correlations
from charm and bottom quarks.  Silicon vertex detector upgrades and higher luminosity
data will greatly help these measurements.  Additionally, these types of measurements will
be very interesting at the Large Hadron Collider.

Preliminary results presented at the conference on extracting the relative contributions to
the electron spectrum from $c$ and $b$ quarks have been superseded by final results, submitted
for publication~\cite{ppg094}.
\bibliography{sickles_wwnd09}
\bibliographystyle{bigsky2009.bst}
 
\vfill\eject
\end{document}